# Bacterial biofilms use chiral branches to escape crowded environments by tracking oxygen gradient

## Authors

Mustafa Basaran[1,2], Tevfik Can Yüce[1], Yusuf Ilker Yaman[1], Roman Vetter[3,4] and Askin Kocabas[1,2,5,6]†

[1]Department of Physics, Koç University, Sarıyer, 34450 Istanbul, Turkey

[2]Bio-Medical Sciences and Engineering Program, Koç University, Sarıyer, 34450 Istanbul, Turkey

[3]Computational Physics for Engineering Materials, ETH Zurich, 8093 Zurich, Switzerland

[4]Current address: Department of Biosystems Science and Engineering, ETH Zurich, 4058 Basel, Switzerland

[5]Koç University Surface Science and Technology Center, Koç University, Sarıyer, 34450 Istanbul, Turkey

[6]Koç University Research Center for Translational Medicine, Koç University, Sarıyer, 34450 Istanbul, Turkey

†Corresponding author: akocabas@ku.edu.tr

# Abstract

Bacterial biofilms collectively develop distinct and ordered structures, including fibers, bundles, and branches. Often, it is unclear how these structural motifs convey specific advantages to bacterial strains under challenging conditions. In oxygen-limited environments, dense bacterial aggregates generally deplete oxygen, which leads to arrest of bacterial growth. However, we observed that biofilm-forming *Bacillus subtilis* could use branching patterns to escape from these crowded regions by tracking the oxygen gradient. The process depends on the chain-forming ability of the biofilm. As a result of collective branching triggered by bending and mechanical buckling, bacteria can extend from the oxygen-depleted biofilm core to the oxygen available periphery. Remarkably, these bacterial branches are strongly chiral and curve in a clockwise direction. Our analysis revealed that the surface friction and axial rotation of the twisting cell wall break left-right symmetry on solid surfaces and drive chiral bending of bacterial chains. We further observed that the chirality of individual branches could propagate across a large scale and shape the macroscopic morphology of the colony under a limited spatiotemporal growth profile. Taken together, our results provide new insights into how simple physical interactions lead to bacterial collaboration and promote the survival of biofilm-forming strains in challenging environments.

# Introduction

Bacterial biofilms are multicellular communities with complex and distinct morphologies[1]. This collective response provides various advantages to bacteria, ranging from antibiotic resistance[2-5] to sliding motility on dry surfaces[1]. A characteristic feature of biofilms is the formation of an extracellular matrix that can attach bacteria together, resulting in a dense colony structure[6]. These complex biofilm structures can have significant impacts on human health by leading to severe infections. At the same time, they also exert essential benefits on plant growth by colonizing roots[7,8]. Understanding the different forms and functions of biofilms is an essential and longstanding goal in microbiology.

*Bacillus subtilis* produces diverse cell types, ranging from solitary bacteria to multicellular biofilms[1,6,9,10]. While natural isolates collectively form biofilms, laboratory strains are generally solitary bacteria[11]. Time-lapse imaging starting from a single bacterium revealed that the early stage of biofilm formation is triggered by the chaining response of bacteria[12,13]. As bacterial chains grow, they undergo successive elastic buckling and form aligned bundles and branches, eventually forming the biofilm structure[13]. Previous studies have shown that a complex interplay of environmental and genetic factors[14-16], including nutrition[17], surface moisture[18], and stochastic expression of regulatory genes[19,20], regulates biofilm formation. Moreover, different aspects of biofilm formation at the air/liquid interface have been linked to impaired respiration and hypoxia[21]. Interestingly, in contrast to laboratory strains which produce flat bacterial layers, wild-type strains produce folded and wrinkled pellicles[9,11,21]. These complex surface interface structures allow nutrients and air to penetrate bulky biofilms more effectively[22,23]. Although these studies reported various essential factors that regulate biofilm formation, very little is known about how specific building blocks of biofilm structures confer fitness advantages to bacteria under challenging environmental conditions.

In the present study, we assessed the benefits of branch formation in growing *B. subtilis* biofilms in oxygen-limited environments. Time-lapse imaging was used to observe spatiotemporal growth dynamics and the formation of specific structural motifs of a biofilm starting from a single bacterium. We found that growing bacterial chains form chiral branching patterns that allow bacteria to track oxygen gradients and escape from oxygen-limited regions.

Our study showed that chaining response provides a strong competitive advantage to *B. subtilis* under confined and oxygen-limited conditions. *Bacillus subtilis* is a root-colonizing organism[7] and confinement and oxygen limitation are common environmental challenges in soil. We suggest that *B. subtilis* may adapt to these environmental conditions by forming biofilms. Similar branching and filamentation patterns are very common geometric forms observed in various organisms ranging from cyanobacteria, and yeast to soil bacteria[24-29] and may share a generic selection advantage across these organisms. Interestingly, on solid surfaces bacterial chains are chiral which can shape overall biofilm morphology. This physical response reflects the intrinsic structural and physiological properties of growing bacteria. We further identified the critical physical interactions that lead to the chirality of the biofilm. Altogether, our results provide new physical insights into a simple form of bacterial collaboration and biofilm growth in complex natural environments.

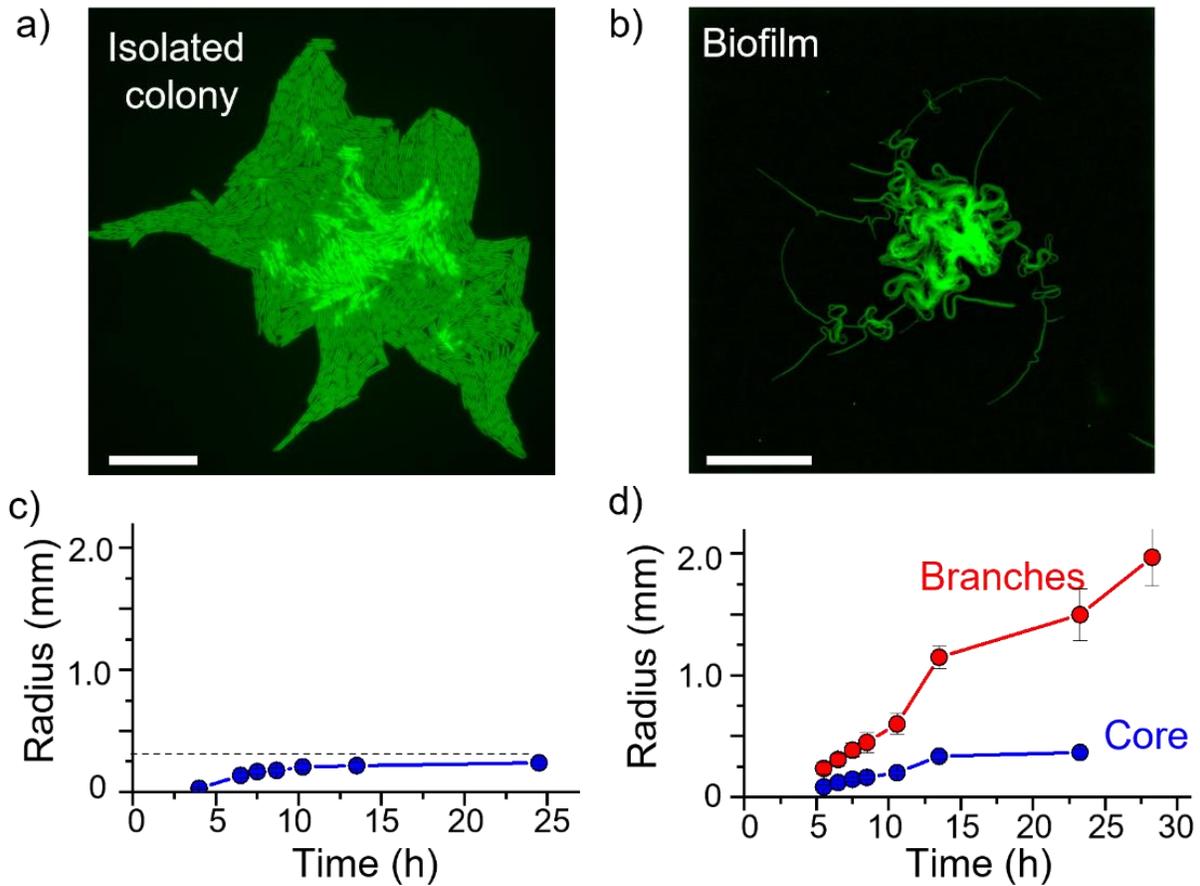

*Figure 1.* ***Experimental observation of chiral branching patterns in growing bacterial biofilms.*** *(a) Fluorescence images of isolated (strain 3610) and chaining bacterial biofilms (strain 168) on a solid LB plate. Isolated bacteria (a) form dense aggregates. However, biofilm-forming bacteria (b) can collectively create branching patterns where the aligned bacteria form a chain. Branching patterns are generally bent in the clockwise direction. (c) The average size of isolated colonies as a function of time. The size of individual colonies is arrested under oxygen limitation. (d) Comparison of the average radius of the branches with the size of the biofilm core. Although the growth of the crowded biofilm core is arrested, the extending branches have a relatively sparse biofilm structure and can continue to grow. Scale bar: 20 μm in (a) and 200 μm in (b). Error bars are defined as standard deviation and averaged over 25 colonies.*

# Results

**Bacterial chains and biofilm formation**

To investigate the spatiotemporal development of biofilm morphology and structural motifs, we used time-lapse fluorescence microscopy. A diluted culture of *B. subtilis* was spread on a flat agar surface to image a single bacterium developing into a colony. Generally, Msgg growth media are used to study biofilm growth however, we noticed that general biofilm forming process is conditional on nutrient-rich Luria broth (LB) plates. We first focused on two strains: a natural isolate (3610) and a laboratory strain 168. In nutrient-limited liquid Msgg, strain 3610 forms biofilms, whereas strain 168 forms a flat colony morphology. The liquid culture-based approach is commonly used for testing biofilm-forming ability because biofilms can collectively colonize the liquid-air interface to gain access to oxygen[21]. However, at an early stage, we observed different responses on nutrient-rich solid LB plates: 3610 formed isolated colonies, while 168 formed chains and complicated biofilm structures (Fig. 1 a and b, Video 1-4). 3610 can only form biofilm at a later stage. Furthermore, we tested other important genes involved in the biofilm formation process, including *slrR, sinI* and *sinR* mutants, and observed that these strains can also develop chains, on the other hand, *ymdB* mutants form isolated colonies on LB plates (Supplementary Fig.1). We should also emphasize that chaining is a very early stage of biofilm formation. This is because the matrix gene *tapA* is stochastically produced (Supplementary Fig.2) at a late stage when bacteria grow on the LB plate. We believe this difference in chain-forming abilities is originating from the interplay of metabolic factors and autolysin processes that controls cell separation[16]. We conclude that chaining behavior is a common response of *B. subtilis* under nutrient-rich conditions and strains 3610 and 168 provide a suitable platform for observing and comparing the specific advantages of the collective chaining ability of bacterial biofilms.

We considered the bacterial chains to be the primary building blocks of biofilm structures. Previously, we showed that the edge of the growing biofilm is unstable because as the biofilm chain grows, accumulated mechanical stress causes localized buckling[13](Video 4). These elastic buckling events successively repeat when the growing chain reaches the critical length and eventually the bacterial chain folds into bundles. At a later stage, growing circular structures appear around the edges. We further observed that the fast-growing central region of the biofilm gradually absorbs the extending branches (Video 3). These experiments identify the main structural motifs and dynamics commonly observed during the early stages of biofilm development. It should be emphasized that these results were observed on a flat agar surface at solid-air interfaces (see Materials and Methods).

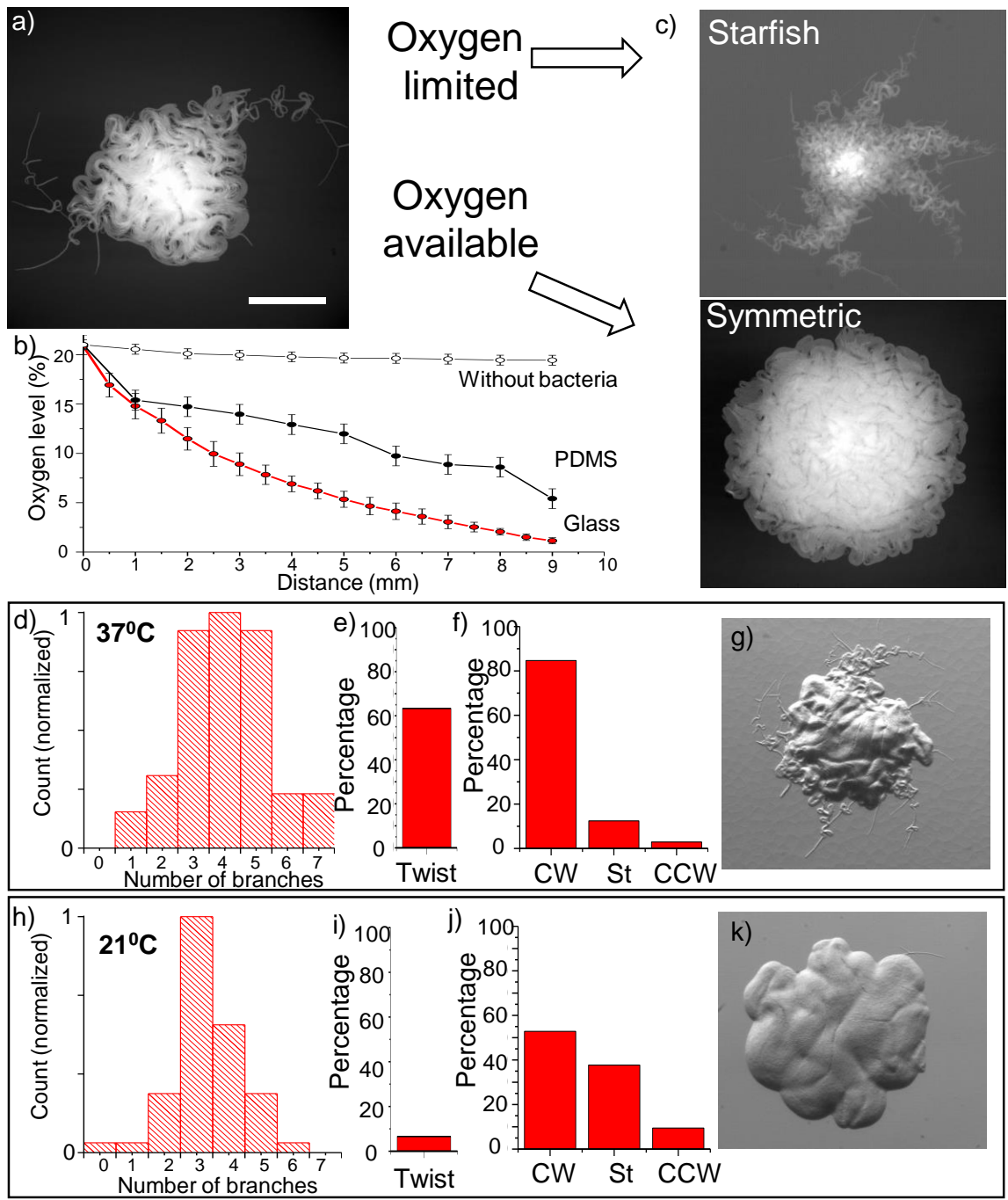

*Figure 2. **Oxygen-dependent biofilm morphology and branching patterns.** Fluorescence image of bacterial biofilms* (strain 168) *growing under different oxygen limitations. (a) Multiple branching patterns are observed at the early stages of colony formation. (b) Experimental measurement of oxygen level [$O_2$] as a function of distance from oxygen providing edge to growing*

*colonies. (c, bottom) Under oxygen available conditions, the fast-growing biofilm core absorbs extending branches and eventually forms a symmetric branchless biofilm edge with circular bundles. (c, top) However, oxygen limitation slows down the growth of the biofilm core and the branching patterns result in the formation of starfish-like colony morphology. (d, h) The impact of growth temperature on the formation of branching patterns. Growth temperature regulates the total number of branches, twists, and bending properties of the branches. Biofilms growing at 37 °C (d) form more branches than those growing at 21 °C (h). (e, i) Twisting macro fiber formation is suppressed at low ambient temperatures. (f, j) Chirality of the branching patterns during biofilm growth. Clockwise-biased chirality is particularly observed at high growth temperatures. Biofilms growing at low ambient temperatures have branchless colony edges. The histograms and percentages are averaged over 50 individual colonies.*

**Bacterial growth and biofilm morphology in oxygen-limited environments**

To further analyze biofilm growth dynamics in an oxygen-limited environment, we limited oxygen availability using a cover glass on the agar surface. Growing colonies obtain necessary oxygen from air-exposed surfaces. However, under glass confinement, oxygen can only diffuse laterally into the colonies. As the colony grows, oxygen depletion results in a radial oxygen gradient, and the oxygen supply to cells on the interior becomes limited. We imaged *B. subtilis* strains growing under these conditions. Initially, on LB agar, the strains 3610 and 168 showed growth rates similar to those observed previously. However, the main difference was noted at later stages (t~15 h), in which colony size of the strain 3610 was arrested (Fig. 1 c and d) as oxygen depletion slows bacterial growth and thereby limits colony size[30,31]. In stark contrast, although the crowded biofilm core of the strain 168 was arrested in size, branches extending from the biofilm continued to grow.

The most parsimonious hypothesis that explains this response is that due to the sparse structure of the branches, self-oxygen depletion is relatively low, and branches can escape oxygen limitations by growing outwardly. To clarify the influence of the confinement, we repeated the experiments using polydimethylsiloxane covers, which is an oxygen-permeable material that provides sufficient oxygen to the colonies under similar vertical confinements (Supplementary Fig. 3). In these experiments, we observed faster radial growth in isolated colonies, which supports the hypothesis that oxygen limitation is critical for isolated colonies. However, in biofilms, oxygen limitations differentially affect the crowded core region and extending branches by defining different spatiotemporal growth profiles. As a control experiment, we coated the glass with thin (50μm) PDMS and observed similar arrested growth (Supplementary Fig. 3) which clarifies the possible impact of the PDMS on bacterial growth. Finally, we independently tested the oxygen limitation by using GFP to RFP conversion upon photobleaching under blue light exposure (Supplementary Fig. 3, Materials and Methods).

We also analyzed the overall colony morphology at a later stage by determining how different growth profiles of the core region and extending branches contribute to biofilm development. Without oxygen limitations, over the course of 5 h of imaging on a regular agar surface, we observed symmetric biofilm morphology with a nematically aligned core structure (Fig. 2a–c). Indeed, multiple extending branches were formed initially but later merged into the fast-growing biofilm core. However, under oxygen limitations, throughout 10 h of imaging, the biofilm developed a starfish-like morphology with multiple arms (Fig. 2c, Video 5), which was the consequence of the slow growth rate of the core region under a radial oxygen gradient. To experimentally confirm the radial oxygen profile, we measured the oxygen concentration by using a fiber optic oxygen sensor (Figure 2b., Materials and Methods). Interestingly, around the vicinity of the glass confined colonies, oxygen level drops to 1%. In contrast, oxygen level under PDMS confinement is above 5%. Moreover, during this period, the biofilm continued to grow through the newly formed branches extending from the arms, and eventually, these fractal-like branches produced a large growing biofilm web (Supplementary Fig.4). Additionally, we observed that branch formation was strongly dependent on growth temperature. Biofilms generated more branches at 37 °C. On the other hand, biofilms typically generated three branches at 21 °C; however, owing to merging, growing biofilms eventually formed circular structures with branchless edges. Other striking features observed were twisting and breakage of bacterial chains during biofilm development[12] (Video 4, 6). Similarly, these events occurred more often at higher temperatures (Fig. 2e). These results suggest that localized buckling, twisting, and chain breakage are three different coupled mechanical processes that can trigger the formation of new branches during biofilm growth (Supplementary Figure 5). We also emphasize that these processes including clockwise (CW) bending (Supplementary Figure 6) occur more frequently at high growth temperatures. As a final control experiment, we imaged freely growing bacteria on flat agar surface. Dense bacterial colonies grew and cover the entire surface very quickly (Video 7). This result indicates that nutrition is not a limiting factor on a free agar surface.

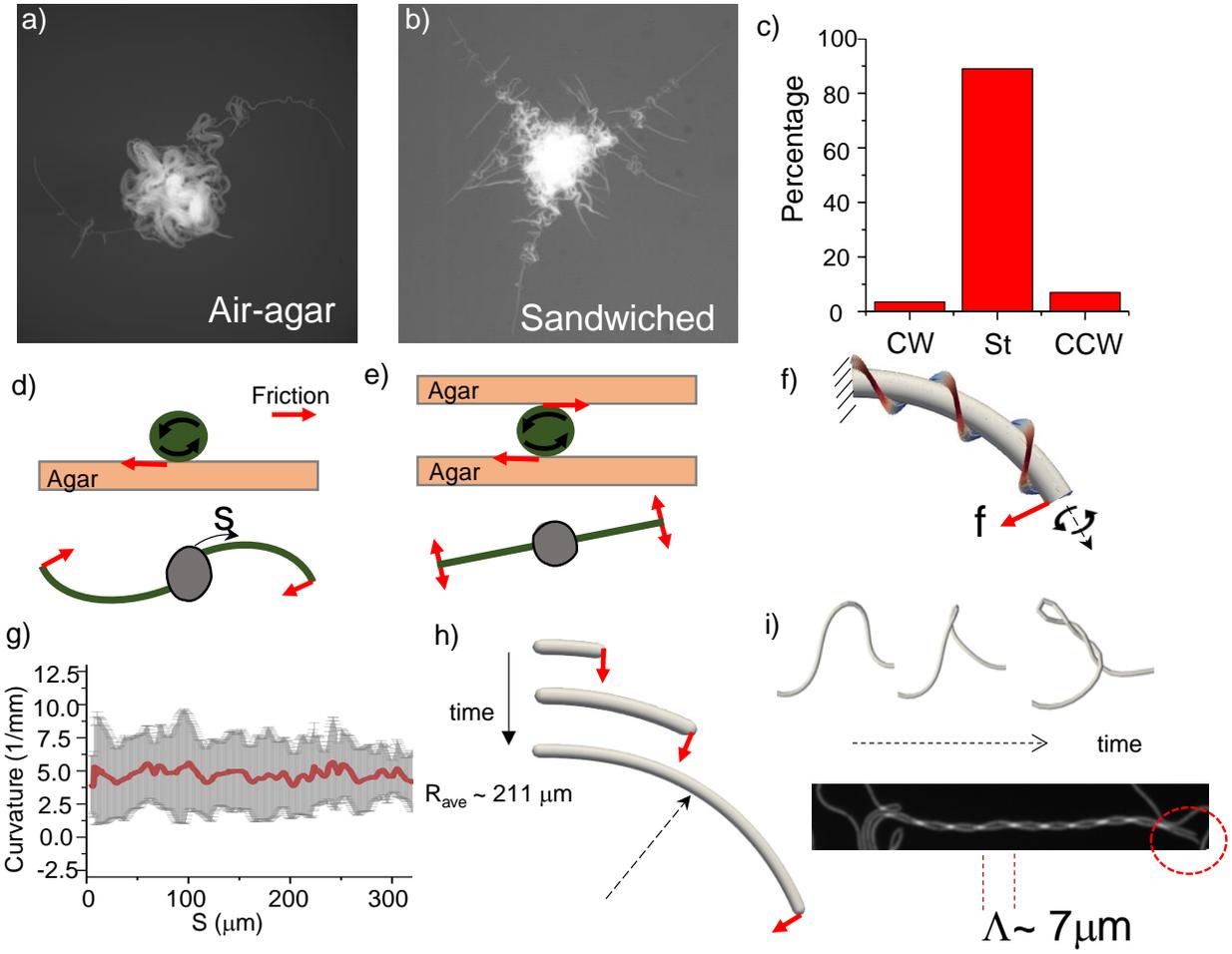

*Figure 3 Chirality of branching patterns. Fluorescence images of growing bacterial biofilm with air-agar and agar-agar sandwiched interfaces. (a, d) B. subtilis chains bend in a clockwise direction at air-agar interfaces. (b, e) At agar-agar sandwiched interfaces, outgrowing branches are generally straight (c). Sandwiched configuration can balance the rotational friction that occurs between the axially rotating cell wall and the agar surface. (f) Schematic representation of B. subtilis cell wall twisting in the right-handed direction during elongation. The twisting cell wall rotates the cell body in the axial direction and experiences rotational friction on a flat solid surface. All together right-handed intrinsic chirality of the bacterial cell wall leads to clockwise bending of the growing bacterial chains (seen from the top). (g) Experimental measurement of*

*bending curvature of the branches shows a constant profile. (h) FEM simulation of the growing rod under constant tip force and surface friction. Growing rods provide similar constant bending curvature. (i) Accumulated stress in a growing elastic rod triggers localized buckling and intrinsic twisting dynamics further leads to the formation of twisting macro fibers. These fibers have a macroscopic twisting periodicity of about Λ = 7 μm and as a result of chain breakage (highlighted in a red circle), they can generate new growing branches.*

**Chirality of growing bacterial branches**

The most notable feature of bacterial branches is their chirality. As branches grow, they gradually bend in a clockwise (CW) direction (imaged directly from the top without any mirror reflection). Hence, we sought to understand why the branches are chiral and what breaks the left-right symmetry in growing biofilms. Although chirality is a common property in biological systems, the detailed physical principles driving symmetry breaking vary and are generally not well understood. To date, various factors, ranging from hydrodynamics[32] to intracellular[33] constraints, have been highlighted as the driving forces for the emergence of chiral responses. Interestingly, the branching-based chiral pattern formation was previously observed in goldfish retina explants[34,35] and fungal hyphae of *Neuropora crassa*[36]. Although there is a large body of literature on the chirality of these systems, the formation of chiral biofilm branches has not yet been studied. Taking into account the findings of these studies, we can naively claim that surface friction is responsible for breaking symmetry. This is because cell wall synthesis in *B. subtilis* has right-handed (RH) structural chirality. However, the rod shape of bacteria growing in liquid media is not intrinsically bent in the preferred direction[37]. It is possible that during growth twisting of the cell wall[38-41] can create axial body rotation. Previous reports from single-molecule imaging of cell wall synthesis provide strong evidence to support this hypothesis[38,41]. It is also clear that the RH chirality of *B.*

*subtilis*[38] cell wall together with the asymmetric surface friction should bend the growing chains in the CW direction. Our experimental observation of CW bending of growing chains strongly supports the friction force-based model. To directly challenge this hypothesis, we sandwiched the growing biofilm between two identical agar surfaces. As expected, under the sandwiched configuration, the friction forces canceled each other out and resulted in the formation of straight bacterial branches (Fig. 3a b,) and eventually triangular colonies (Supplementary Figure 7). On the other hand, asymmetric sandwiching between glass and agar interfaces resulted in similar CW bending (Fig.2b). From these observations, we concluded that chiral bending originates from the biophysical interactions between the bacterial cell wall and the agar surface.

We have to emphasize that different forms of macroscopic chirality have been previously linked to the intrinsic cell wall structure. It has been reported that growing colonies of mixed *Escherichia coli* populations give rise to chiral segmentation of the colonies[42], interestingly this is associated with direct fitness advantages[43]. Cell wall chirality and filamentation of the bacteria were also related to this process[44]. We believe that this chiral segmentation is associated with similar surface interactions, but is influenced by more complex biomechanical factors, including bacterial alignment and cell-to-cell adhesion in *E. coli* colonies. Moreover, the cell wall chirality of *B. subtilis* has received significant attention owing to the formation of twisted macrofibers. Biofilm-forming *B. subtilis* in liquid culture produce large-scale twisted macrofibers[45-47], which are generally RH. We believe that all these chiral structures are intricately related. In the present study, we observed a direct projection of cell wall chirality onto the bending dynamics of branching biofilm patterns.

To gain further insight into the relationship between cell wall synthesis and chiral bending, we attempted to manipulate the cell wall structure. Recent studies[48] have reported powerful tools to manipulate cell wall machinery and we hypothesized that this could be used to manipulate chiral bending. Two distinct pathways control the directional (MreB-guided) and random insertion of glycan into the cell wall, which has a direct impact on the intrinsic twist of the cell wall. These insertion mechanisms can be independently controlled by chemical induction using xylose and isopropyl thiogalactopyranoside (IPTG). To observe the impact of directional insertion, which provides a more twisted cell wall structure, on chiral bending, we imaged bacteria growing under a wide range of inducer concentrations. Overexpression of MreB resulted in kinkier bacterial chains instead of strong chiral bending; on the other hand, the more random insertion of glycan resulted in rounded cells and diminished chiral bending (Supplementary Figure 8). We noted that precise manipulation of chirality is not easily achievable on agar plates. However, during these experiments, we observed that at some inducer concentrations, bacterial colonies were macroscopically chiral (Supplementary Figure 9). It should be noted that large-scale chiral morphology was observed only under oxygen limitations. We further hypothesized that this unexpected colony morphology, although originating from different limitations, creates similar spatiotemporal growth profiles leading to formation of starfish like colony. Dion et al.[48] also reported that controlling cell wall synthesis has an additional impact on the growth rate; particularly, suppression of MreB expression using low dose xylose induction slows down bacterial growth[48]. It is likely that limited xylose diffusion from agar surfaces into a growing colony, similarly, slows down the growth rate at the center of the colony, as observed under radial oxygen limitation. Motivated by this idea, we further challenged this scenario by reproducing a similar macroscopic chiral colony morphology under nutrient limitation by decreasing the nutrient content of the LB plate (Materials and Methods) and decreasing the agar thickness to 1 mm to further enhance this limitation. As expected, under these limitations, growing bacterial biofilms gave rise to similar macroscopically chiral colonies without any confinement and oxygen limitation (Supplementary Figure 10). These experiments suggest that the limited spatiotemporal growth rate of a colony is a generic mechanism that controls biofilm morphology at large scale. Thus, colony chirality could be a macroscopic indicator of the intrinsic limitations of colony physiology.

**Elasticity of bacterial branches**

We then focused our attention on the detailed shapes of the chiral branches. Different spiral forms, Euler and natural types have been observed in branching fungi[36] and segregated bacterial colonies[42,44], respectively. To characterize the bending form of the observed biofilm branches, we traced and quantified the curvature of the individual filaments extending from the biofilm core and observed varying profiles (Supplementary Figure 11,12). While some branches deviated from the initial straight growth to a circular form, some were initially bent, but become straight at a later stage. However, the averaged trajectories show a constant mean radius of growth curvature ($\kappa = 1/R$), which is approximately $R = 211$ μm (Fig. 3g). Another interesting feature is that, as the chain grows, instead of sliding on the surface, the entire chain follows the same trajectory defined by the tip. In order to gain more understanding of the mechanical properties, we performed a finite element analysis of the growing elastic rods under different configurations. We modified recently developed finite element method (FEM, Materials and Methods) codes to include friction-induced torque, plasticity, and intrinsic twist. After extensive simulation, we noticed two generic final filament configurations, strongly buckled and sliding of the entire filament. After testing various types of surface forces, we found that the constant curvature of the growing rod could be captured by applying a follower force to the tip[49] (Video 8). The localization of the force to the tip is reasonable because the chain nature of the branches may relax the axial rotation across the filament and torque may be more effective only around the tip region. It should be emphasized that different scenarios are also possible during growth, such as agar surface deformation and adhesion to the surface due to bacterial nanotubes. In particular, nanotube structures form a web around bacteria[50], which can fix growing bacteria to the surface, allowing only the tip to rotate and bend. We further observed that the intrinsic twist of the growing rod could also lead to the formation of twisted macro fibers (Fig.3i), which are generally triggered by localized buckling that occurs around the center of the growing chains. We further referred to the existing literature on the dynamics of this macro fiber formation[45].

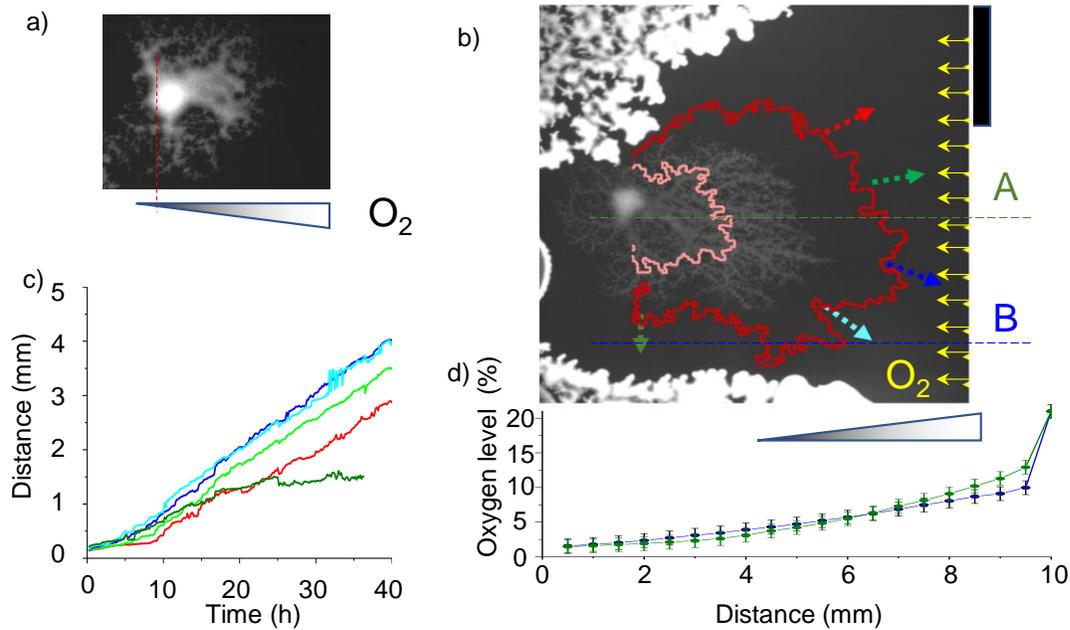

*Figure 4 Branching bacterial biofilms can track oxygen gradients. (a) Fluorescence image of an individual bacterial colony growing around the edge of the inoculation region. The colony shows asymmetric morphology towards the oxygen available region. Redline represents the initial position of the single bacterium. (b) Spatial competition of isolated and biofilm-forming bacteria. The growth of isolated bacterial colonies is arrested under oxygen limitation; however, biofilm-forming bacteria can grow toward the oxygen available region. (c) Interfacial growth of the*

*branching pattern in different directions as a function of time. A constant growth profile towards oxygen available region and sparse web-like biofilm structures supports the oxygen-limited growth of the branching patterns. (d) Oxygen level as a function of distance from the colony to the oxygen providing edge across two different cross sections labeled as A and B on the image, Scale bar: 1 mm.*

**Biofilms can track oxygen gradient**

Finally, we observed that the colony morphology of confined and oxygen limited bacterial biofilms growing around the edge of the inoculation area was asymmetric; entire bacterial biofilms grew outwardly (Fig.4a). It is possible that the colonies can grow more effectively towards the oxygen available region. This response could represent an additional ability of biofilms to track oxygen gradients. To test this passive oxygen tracking strategy, we analyzed a single biofilm-forming bacterium under a sharp oxygen gradient profile. It is very challenging to create gradients on wet agar surfaces; thus, we used U-shaped and patterned isolated bacterial colonies (3610) as oxygen sinks to define varying oxygen concentrations. As expected, the dense bacterial region depleted oxygen and arrested bacterial growth; however, oxygen can diffuse laterally from the open end of the U-shaped region. Figure 4b, c shows the typical response of a single biofilm-forming bacterium in an oxygen gradient. Biofilms preferentially grew toward the oxygen available region (Fig.4 c, Video 9) by forming branching patterns. We experimentally measured the oxygen gradient across the U-shaped region by using fiber optic sensor (Fig. 4d). Although initially there is only one biofilm-forming bacterium on the surface, due to its branching ability it can eventually invade the entire plate. Notably, after reaching the edge, bacteria could switch to a motile state and swim across the edge of the cover (Video 10), as owing to the capillary effect, the liquid is concentrated around the sharp edge of the cover. These results further suggest that growth-induced oxygen tracking ability is an effective strategy that bacteria utilize to spread and grow in crowded and oxygen-limited environments and that switching between chaining and motile states[19,20] is also critical to facilitating the invasion of bacterial strains into large wet areas.

Finally, we hypothesize that filamentation patterns could be very geometric forms in nature. In order to test this idea, we randomly sample the soil in the forest near our campus. Surprisingly, in our first trail we identified 3 different branch forming bacteria and one chiral fungus (Supplementary Figure 13) on our nutrition limited plates where we have similar arrested growth and escaping biofilms by branching. Two of them show similar branching phenotype we observed using *B. subtilis*. Among those the most notable one is the biofilm forming bacteria with chiral branches growing at room temperature (Supplementary Figure 14). This is significantly different response because *B. subtilis* only form chiral branches at growth temperature. Taken together all these observations supports the idea that bacterial filamentation and branching is a generic form in challenging natural environments.

## Discussion

Bacterial biofilms are the basic structures that facilitate collective behavior and bacterial cooperation. This behavior is associated with various structural complexities that provide competitive advantages. Floating pellicles that colonize the air-liquid interface upon exposure to hypoxic conditions are the most notable form of biofilms. This behavior is generic and has been extensively studied in *Pseudomonas aeruginosa* and *B. subtilis*. Floating pellicles have complicated structural morphologies, including aligned filamentous bacteria and extracellular polysaccharide matrix. It has previously been reported that natural bacterial isolates of *B. subtilis*, not laboratory strains, show this form of biofilm morphology on air-liquid interfaces. In the present study, we report a very early form of biofilm formation, characterized by only chaining and branching behavior. Even without extracellular polysaccharides, this basic collective response provides a strong advantage allowing bacteria to escape from crowded regions. More notably, we identified the critical physical interactions that lead to the chirality of the branches. Although the specific advantage of the chiral shape over the straight shape of the branches is not known, it is evident that chirality reflects the intrinsic structural properties of the bacterial cell wall and its ability to generate new branches. Chiral branches are more susceptible to chain breakage and twisted macro fibers, which further promotes the formation of additional branches. Furthermore, macroscopic chiral colony morphology also reflects the limited spatiotemporal growth profile of the colony. This morphological observation may help in identifying specific types of bacteria and their inner physiological state with the naked eye. We also reported the oxygen tracking ability of biofilms based on their branching behavior. It is known that the genetic switch between motile and chaining states is stochastic. We observed that biofilms could finally switch to the motile swimming state upon reaching the oxygen and moisture available regions. All these results support that different building blocks of biofilms provide various advantages facilitating bacterial spread. Our findings also raise new questions, such as the mechanism through which biofilms override surface tension during branch formation, as particularly on wet surfaces, elastic-capillary effects could trigger crumpled forms, making it is challenging for growing biofilms to produce outgrowing branches. We speculate that bacterial nanotubes[50] play a significant role in this process. Future studies are required to investigate the contribution of these effects. Altogether our findings are of potential relevance for understanding the importance of bacterial filamentation, stability of dense bacterial populations and may shed light on the progression of infectious diseases in complex environments.

## Materials and Methods

**Bacterial preparation and growth conditions.** Bacterial cultures were grown in Luria-Bertani (LB) broth at 37 °C on a shaker. An overnight culture was diluted 100x and grown for 8 h. The culture was diluted 10000x, and 10 µl of culture was seeded on a 1.5 % LB solid agar plate. In order to form a crowded environment, the inoculation area was tuned to achieve 4-5 bacteria /mm$^2$. Independently, low salt, 1.5 % Lennox LB agar (Conda) was also used to test the chain-forming ability and isolated colony formation. 1-day-old LB plates were used for time-lapse imaging. 100 µm thick glass and 0.5 mm thick PDMS cover were used for vertical confinement. As a control experiment, 50 µm thick PDMS layer was coated on thin glass by using spin coating technique and then cured at 50 $^0$C for 1 day. For nutrient limitation the concentration of the yeast extract in LB plates was decreased to 20% of regular concentration. Nutrition limited plate was also used for environmental sampling.

The background strain TMN1138 was obtained from R. Losick Lab and bMD620 was obtained from E. C. Garner Lab. All other strains given in the list were received from Bacillus Genetic Stock Center.

**Microscopy imaging.** Fluorescence time-lapse imaging was performed using Nikon inverted and Stereo SMZ18 microscopes. Images were obtained using Andor EMCCD camera. Time intervals between successive images were set to 5-15 minutes.

**Oxygen Measurement**

Oxygen level around growing bacterial colonies was measured by using a florescence based fiber optic oxygen sensor (PreSens, Microx TX3). During [O2] measurements the sensor probe was precisely inserted between cover and agar surface by using a motorized stage (Video 11). The fiber optic sensor has a polymer coating and O$_2$ can quench the fluorescence signal. Oxygen limitation under glass confinement was independently confirmed with oxygen dependent photobleaching. GFP to RFP conversion strongly depends on oxygen level[51]. 10 min blue light exposure (I~1mW/mm$^2$) was used to observe GFP-RFP conversion and oxygen limitations (Supplementary Fig. 3).

**FEM Simulation.**

In order to simulate elastic bending of growing biofilm, we employed a parallel finite element library written in C++. The details of the algorithm were previously published[13,52]. The main modification of the current version is the implement of the tip force. Analogous to, the bacteria were modeled as an isotropic, linearly elastic continuum whose initial stress-free shapes were spherocylinders. The bacteria were assumed to maintain a uniform circular cross-section with radius $r = 0.5$μm, a mass density of $1$gcm$^{-3}$, a Young's modulus of $E = 5300$Pa, and a Poisson ratio of $\nu = ⅓$. Under varios surface friction and axial torgue applied on the growing rod we observed two caracteristics; sliding of entire rod on the surface or strong elastic buckling. In order to maintain the constant curvature, we optimized the surface friction and localized the external force to the tip of the growing rod.

**Author contributions**

M.B, T.C.Y., Y.I.Y, and A.K. designed and performed experiments, analyzed data, and developed the imaging systems. R.V. designed and developed the FEM simulation toolbox. MB. and A.K. prepared the draft, and all authors contributed to the final writing of the manuscript.

## Acknowledgments

This work was supported by Tübitak, Project No: 121C159. We thank Menderes Işkın and Livio. N. Carenza for critical reading of the manuscript. We thank Zeiss, Turkey division and Kenan Doğru for digital microscopy training and installation.

## Competing interests

Authors declare no competing interests.

## Tables

| Strain | Parent | Operation | Genotype |
|---|---|---|---|
| BAK47 | 168 | Transformed with plasmid ECE321 from Bacillus Genetic Stock Center | *amyE:*:Pveg-sfGFP (Spc) |
| BAK49 | TMN1138 | | amyE:: PtapA-cfp (Cm) sacA:: Phag-mKate2L (Kan) hagA233V (Phleo) |
| BAK60 | SinI 1S98 | | sinI::kan-50 |
| BAK61 | SinR 1S97 | | sinR::phl |
| BAK62 | 3610 | Transformed with plasmid ECE321 from Bacillus Genetic Stock Center | *amyE:*:Pveg-sfGFP (Spc) |
| BAK73 | bMD620 (derivative of PY79) | | amyE::erm Pxyl-mreBCD, ΔmreBCD::spc, yhdH::cat Pspank-ponA, ΔponA::kan |
| BAK115 | TMN1138 | Transformed with plasmid ECE327 from Bacillus Genetic Stock Center | *amyE:*:Pveg-mKate (Spc) sacA:: Phag-mKate2L (Kan) hagA233V (Phleo) |
| BAK129 | YmdB (BKK16970) | | ΔymdB::kan |
| BAK130 | SlrR (BKK34380) | | ΔSlrR::Kan |
| BAK131 | Isolated from soil | | |

*Table 1: List of strains used in this study.*

# Video Captions

**Video 1** Time-lapse imaging of growing isolated bacterial colony (strain 3610) on a LB plate. The video shows the OCC image of a growing colony. Video is associated with Figure 1a.

**Video 2** Time-lapse imaging of growing isolated bacterial colony (strain 3610). Video shows the fluorescence image of the colony. The video indicates the isolated form of growing and dividing bacteria forming a multilayered dense structure.

**Video 3** Time-lapse imaging of growing bacterial biofilm (strain 168). Video shows the OCC image of chaining bacteria. Video is associated with Figure 1b.

**Video 4** Time-lapse imaging of bacterial biofilm (strain 168) growing at 37 $^0$C. Video shows the fluorescence image of chaining bacteria. The video indicates the localized buckling events, chain breakage and, the twisting dynamics of macrofiler formation.

**Video 5** Time-lapse imaging of bacterial biofilm (strain 168) growing at 37 $^0$C under glass confinement. Due to oxygen limitations, the growth of the biofilm core is arrested. The extended branches can grow and leads to the formation of starfish-like biofilm morphology. The Video is associated with Fig.2c.

**Video 6** Dynamics of twisting macro fibers. Time-lapse imaging of bacterial biofilm (strain 168) growing at 37 $^0$C. The Video shows OCC image of growing twisted macro fibers around the edge. The Video is associated with Supplementary Figure 5e . Growing twisting macro fiber can be observed around bottom left corner of the biofilm. OCC image enhances the visibility the periodic structure of the macrofiber.

**Video 7** Time-lapse imaging of bacterial biofilm (strain 3610) freely growing on agar surface at 21 $^0$C. This video indicates that crowded bacterial population is not nutrition limited on regular LB plates.

**Video 8** Finite element method simulation of growing elastic rod under constant follower tip force. Surface friction keeps bacteria to grow along the same curved trajectory.

**Video 9** Branching biofilm grows towards oxygen available region. Oxygen gradient is defined by U-shaped crowded isolated bacteria and single biofilm-forming bacterium grow towards oxygen available region. In horizontal direction the growth (dark green arrow) stops before touching the boundary.

**Video 10** Switching from chaining to motile state. After reaching the edge of the glass cover, bacteria (strain168, BAK47) switch to a motile state and quickly spread across the edge where the liquid is concentrated due to the capillary effect. The increase in the brightness is originating from the activity of pveg promoter driving GFP expression in a liquid environment.

**Video 11** Experimental measurement of oxygen concentration by using fiber optic sensor. The Video is associated with Fig.2b.

**Supplementary Figures:**

Chaining

Isolated

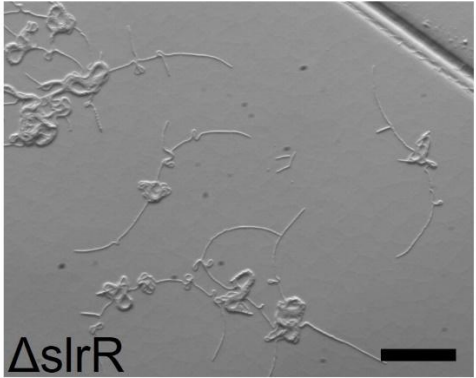
ΔslrR

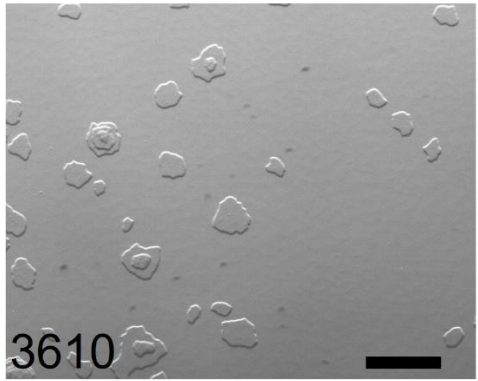
3610

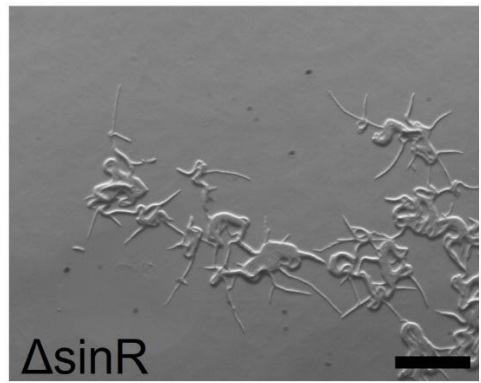
ΔsinR

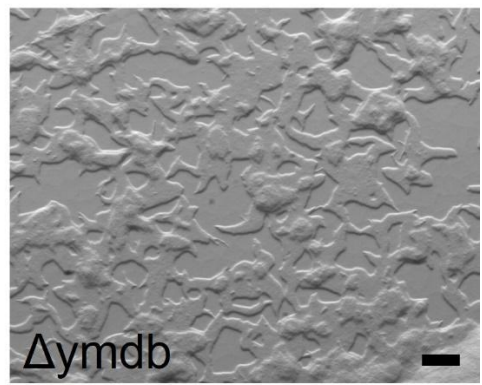
Δymdb

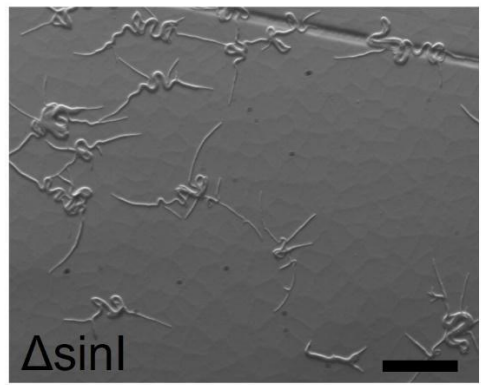
ΔsinI

***Supplementary Fig. 1*** *Various bacterial strains grown on LB plate exhibiting chaining and isolated bacterial colonies.*

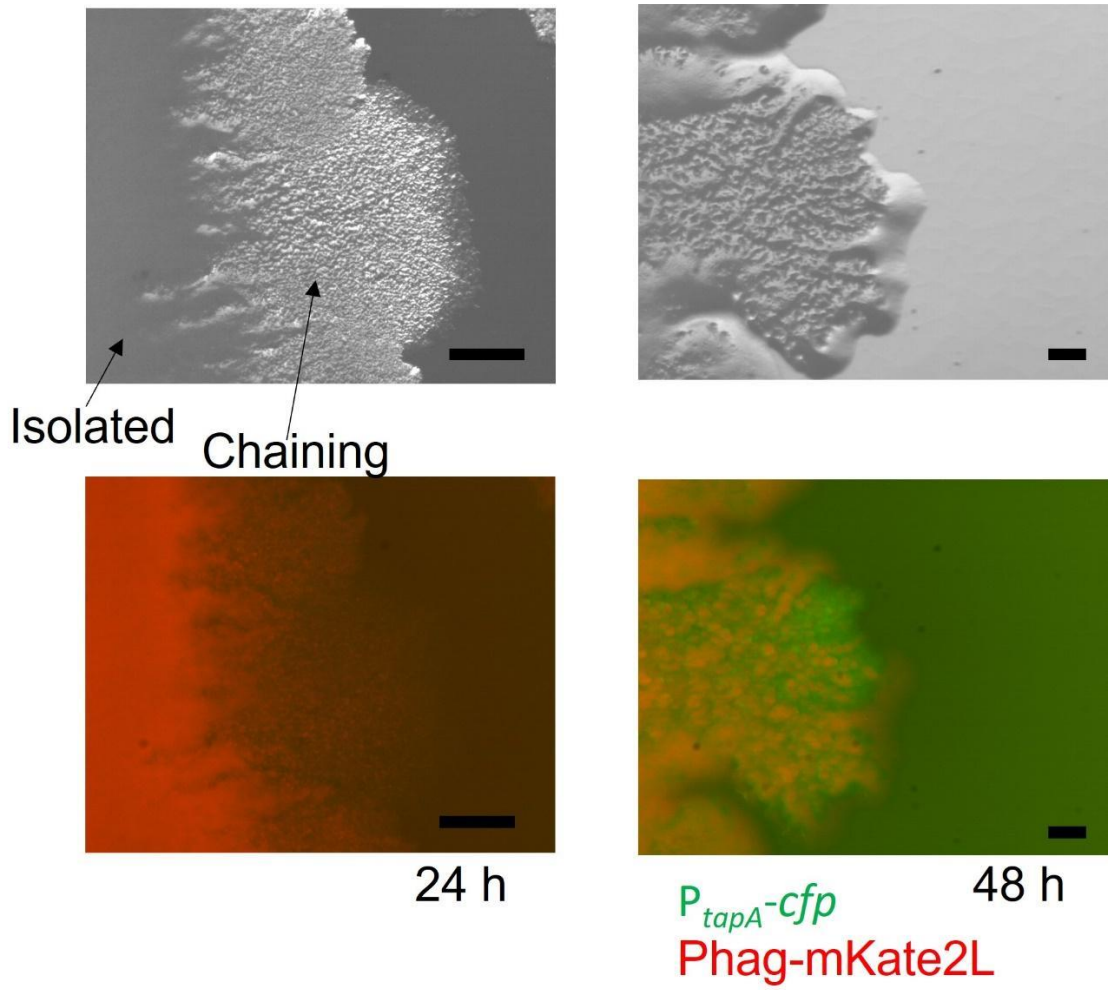

***Supplementary Fig. 2*** *Florescence image of* P*tapA*-*mKate2 expression for a matrix reporter.* TapA is stochastically expressed at late stage after chaining stage observed around the edge of the biofilm.

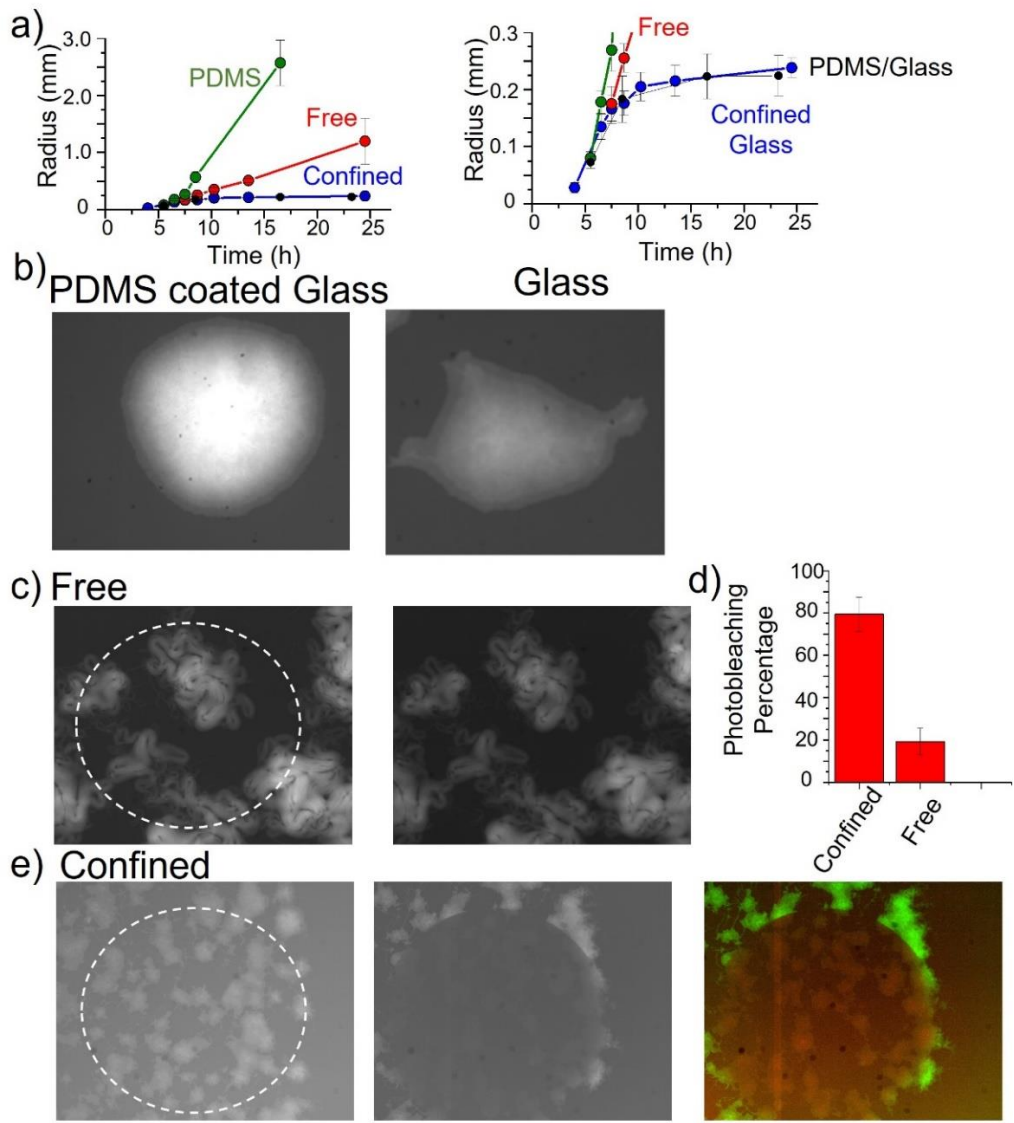

***Supplementary Fig. 3*** a) *Comparison of the average radius of growing isolated bacterial colonies under thin PDMS and PDMS coated glass confinement with the size of the freely growing colonies and glass confinement. PDMS cover can provide sufficient oxygen to penetrate the colony. Glass and PDMS coated glass block oxygen diffusion and arrest the growth. As a result, the colony grows fast in the confined 2D region. b) sample colony images growing under PDMS coated glass and only glass cover. Colonies are generally rounded under PDMS cover but presence of PDMS does not change the arrested growth profile c) Oxygen limitation was confirmed with GFP to RFP conversion upon photobleaching. White circle indicates the exposed region. d) Under oxygen limited confined environment color conversion is significantly higher than that of the colonies freely growing on agar surface. e) Florescence image of GFP RFP and superposition of both colors indicating the oxygen limitation.*

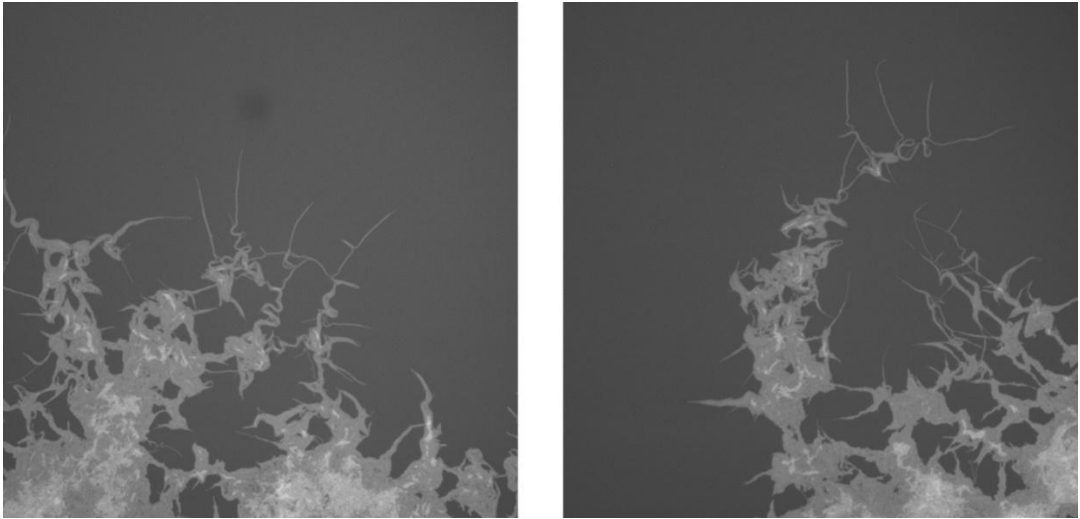

***Supplementary Fig. 4*** *Sample images of web like biofilm structures which are formed at late stage.*

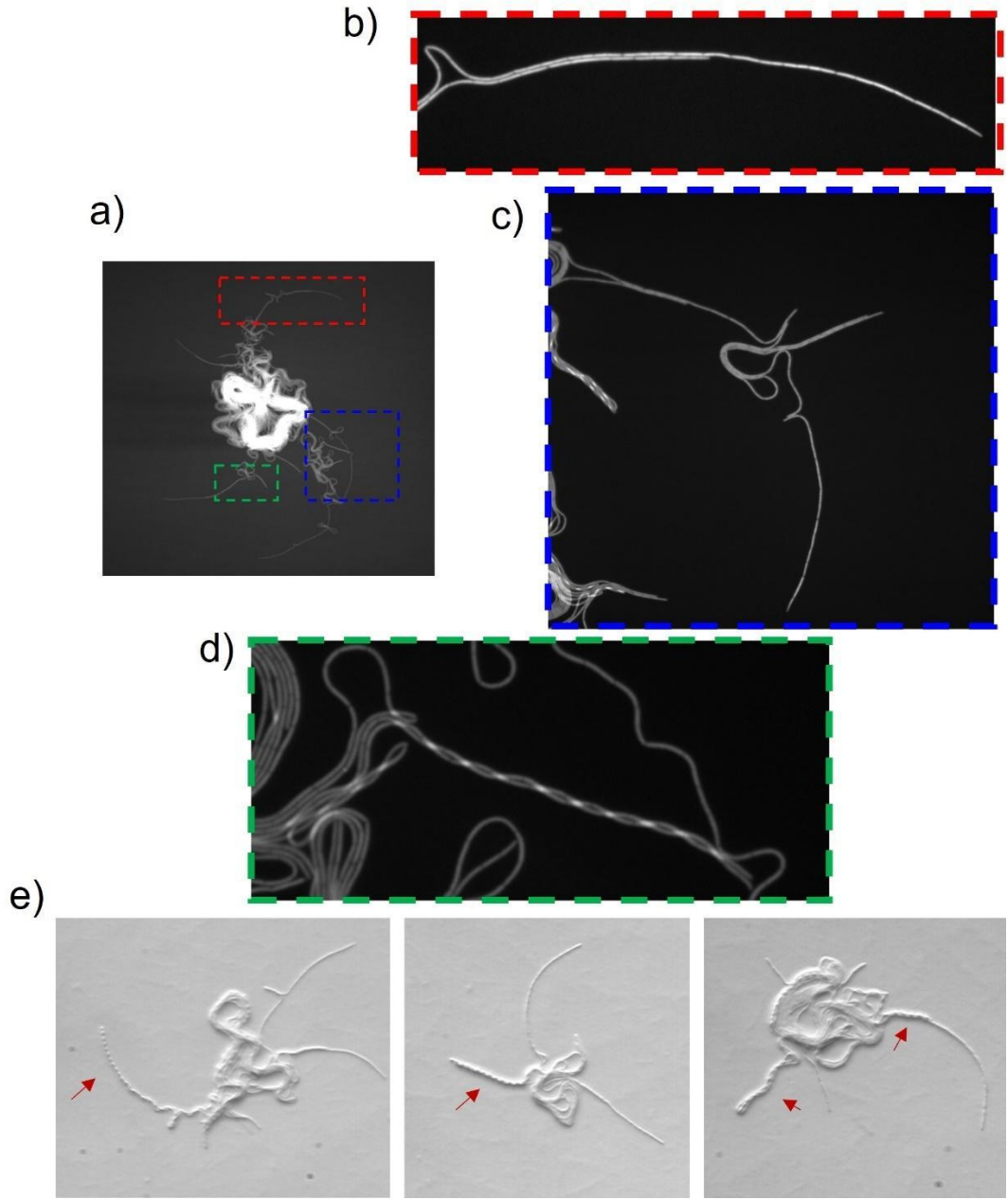

***Supplementary Fig. 5*** *a-d) Fluorescence image of various forms of chaining bacteria. Double-layer, folded, and twisted chains are labeled around the biofilm. e) OCC image of twisted growing macro fibers (red arrows). Twisted fibers (Video 6, bottom left corner) can be observed more often around the biofilm growing at $37^0C$.*

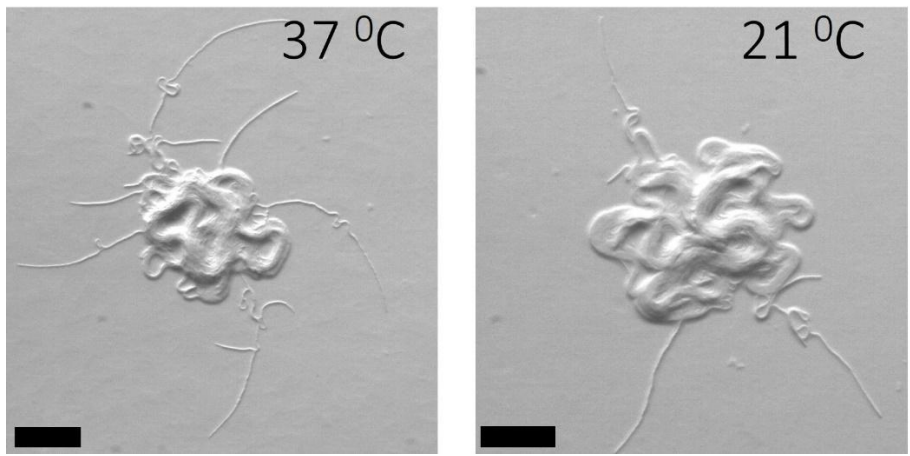

***Supplementary Fig. 6*** *The effect of growth temperature on the chirality of the bacterial branches. Biofilm growing at $37^0C$ (left) forms more chiral branches. scale par is 100μm.*

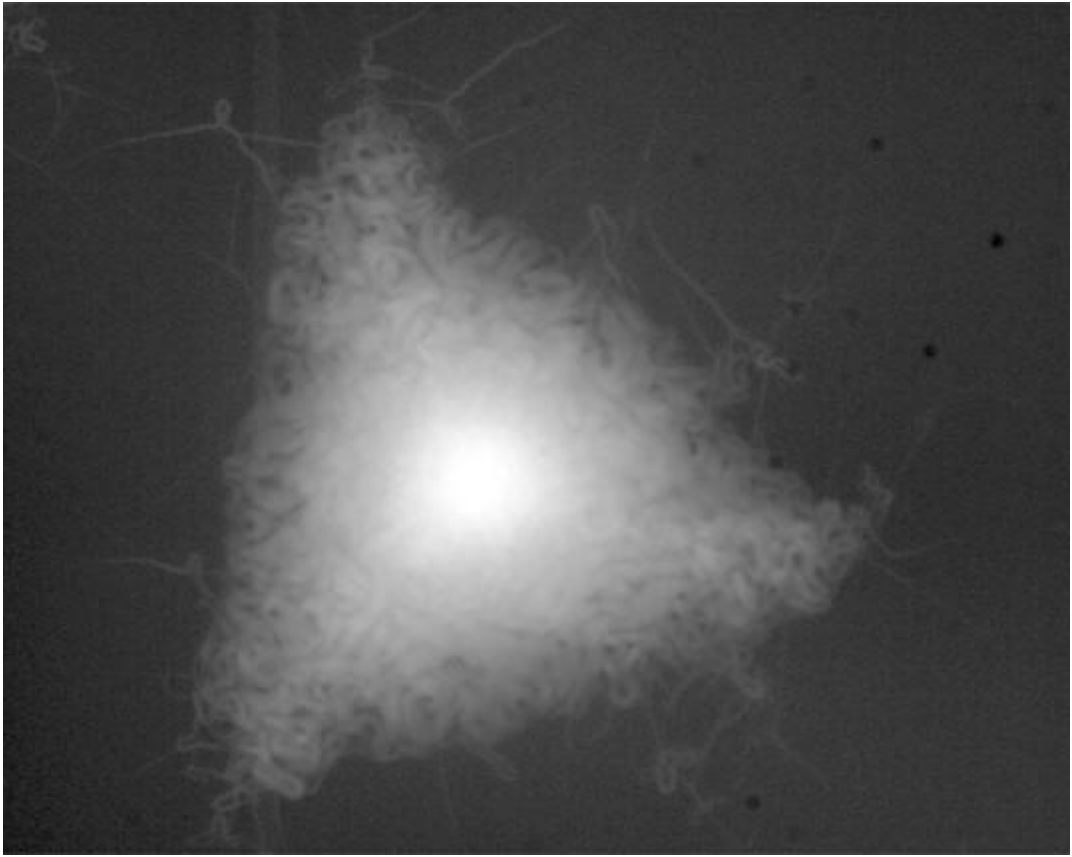

***Supplementary Fig. 7*** *Sandwiched bacterial biofilm forms straight chains which eventually leads to the formation of triangular colony morphology.*

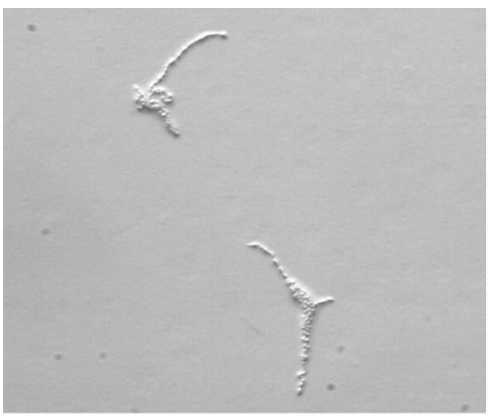
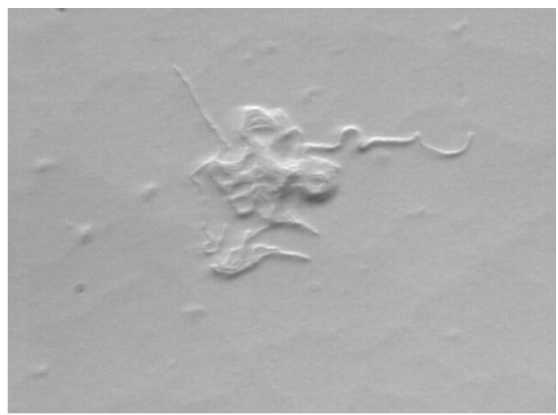

0.5 mM Xylose        50 mM Xylose

***Supplementary Fig. 8*** *Low MreB expression (0.5mM xylose) form rounded bacteria and high MreB (50 mM Xylose) drive skinny and kinkeer chaining biofilm.*

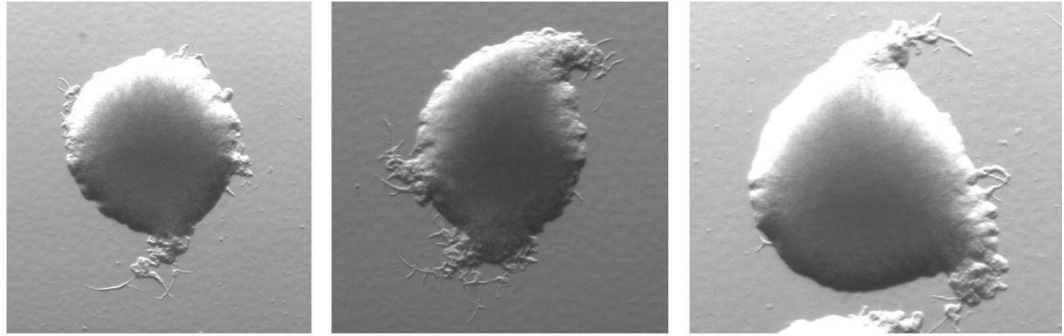
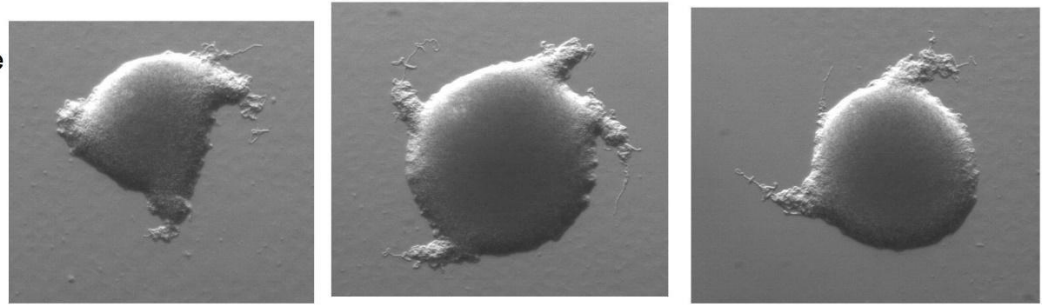
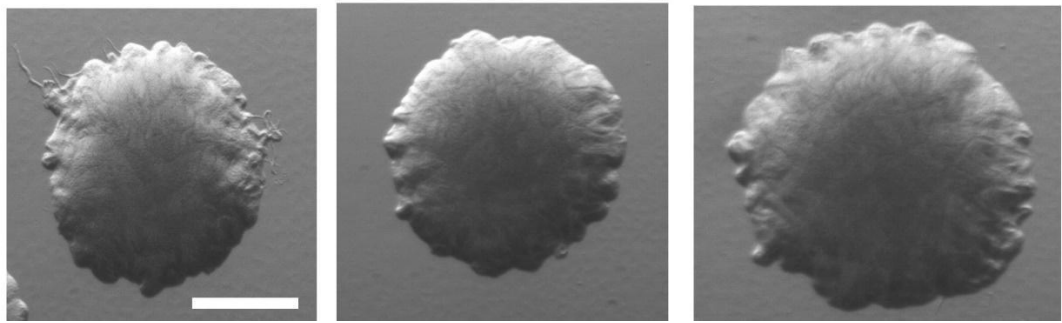

***Supplementary Fig.*** *9 Biofilms (strain bMD620) growing on LB plate with various inducer concentrations. 5mM Xylose leads to the formation of macroscopic chiral morphology.*

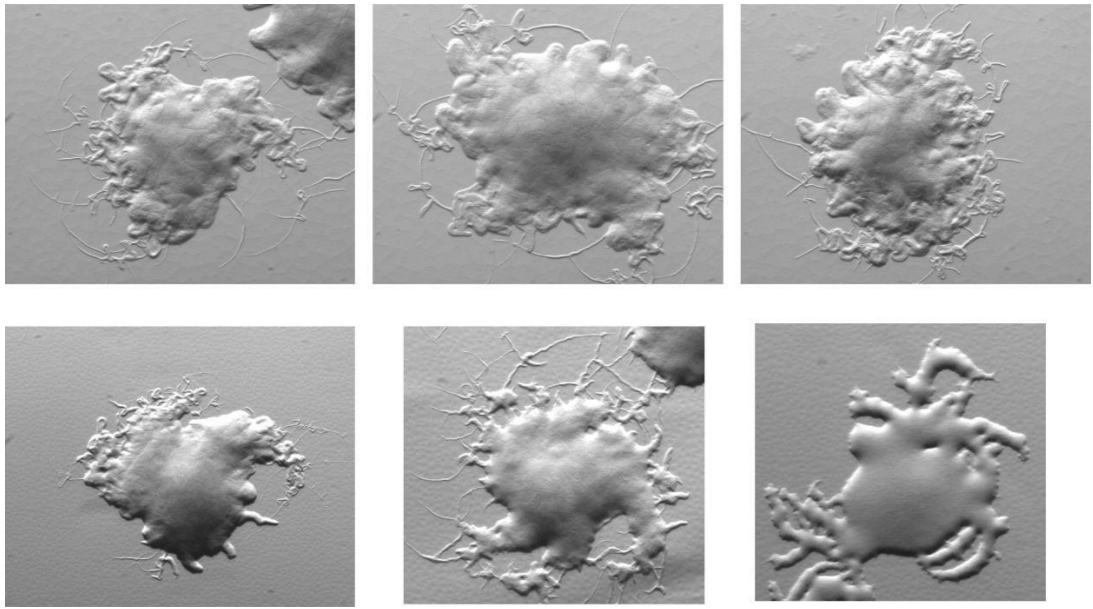

***Supplementary Fig. 10*** *Chiral biofilm morphology (strain 168) induced by nutrient limitation on thin LB plates.*

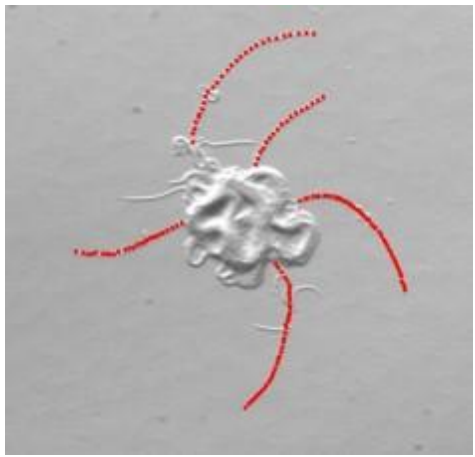

***Supplementary Fig. 11*** *Sample image of traced bacterial branches. Chiral branches were digitized to quantify the curvature.*

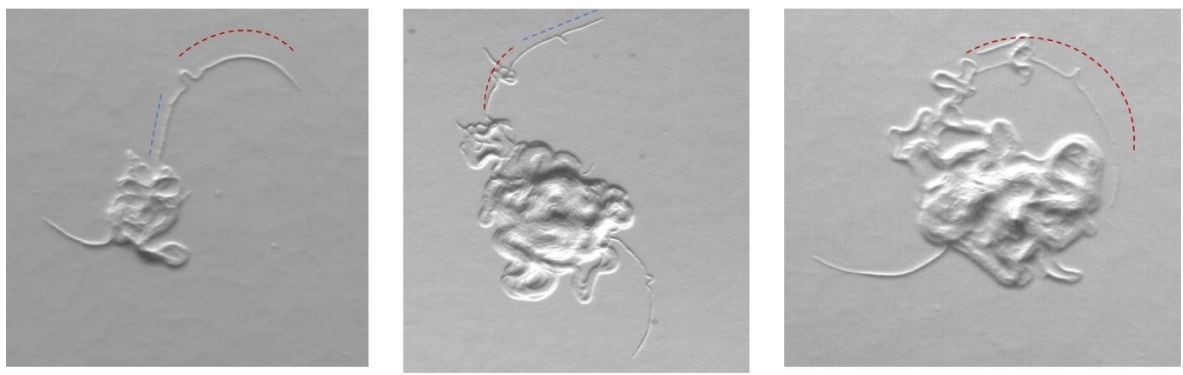

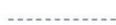 Straight   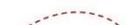 Curved

***Supplementary Fig. 12*** *Sample images of chiral branches with geometric variations.*

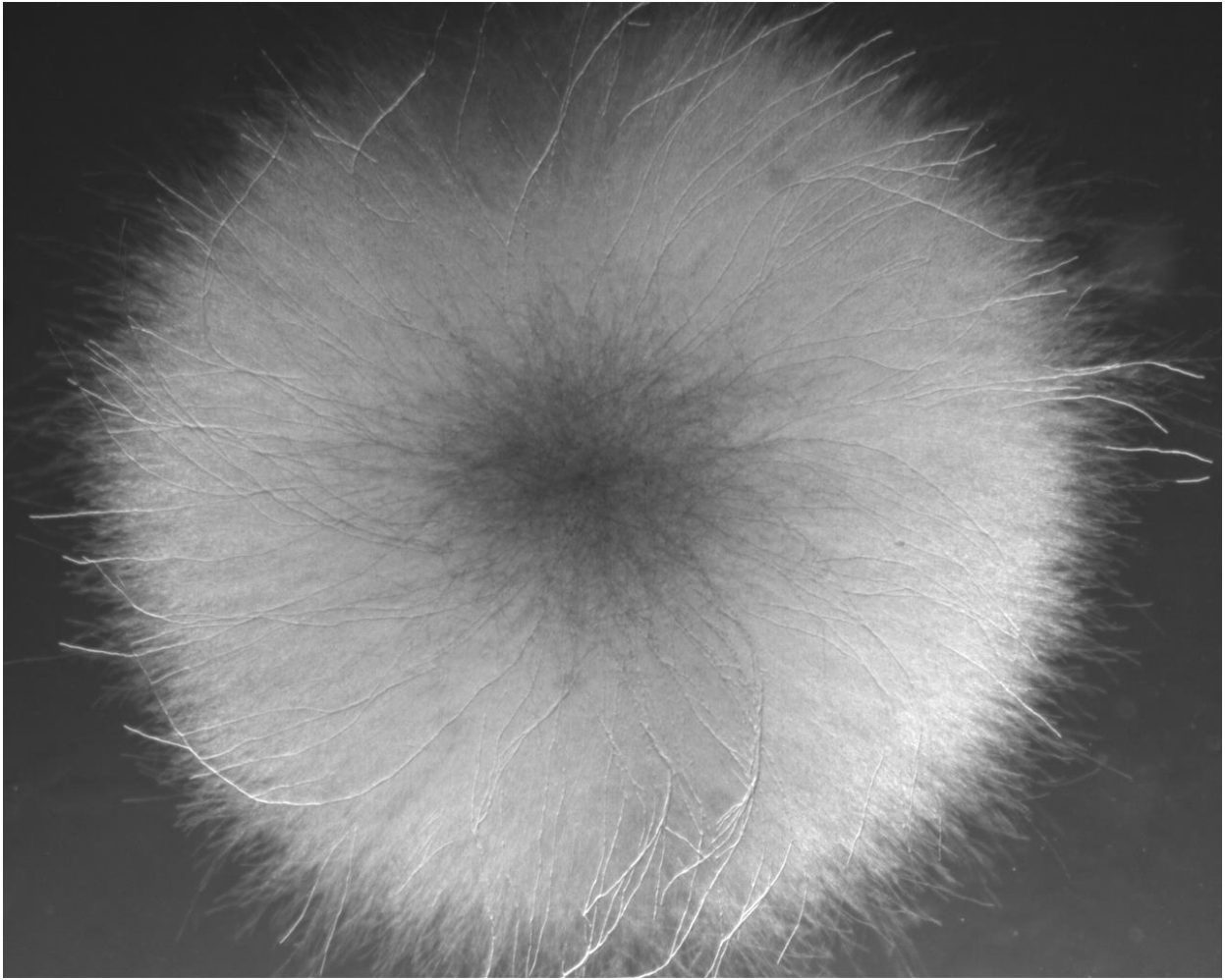

***Supplementary Fig. 13*** Image of chiral fungi isolated from the soil. Branches growing on the surface show similar CW rotation however the branches growing in the agar has radially straight profile.

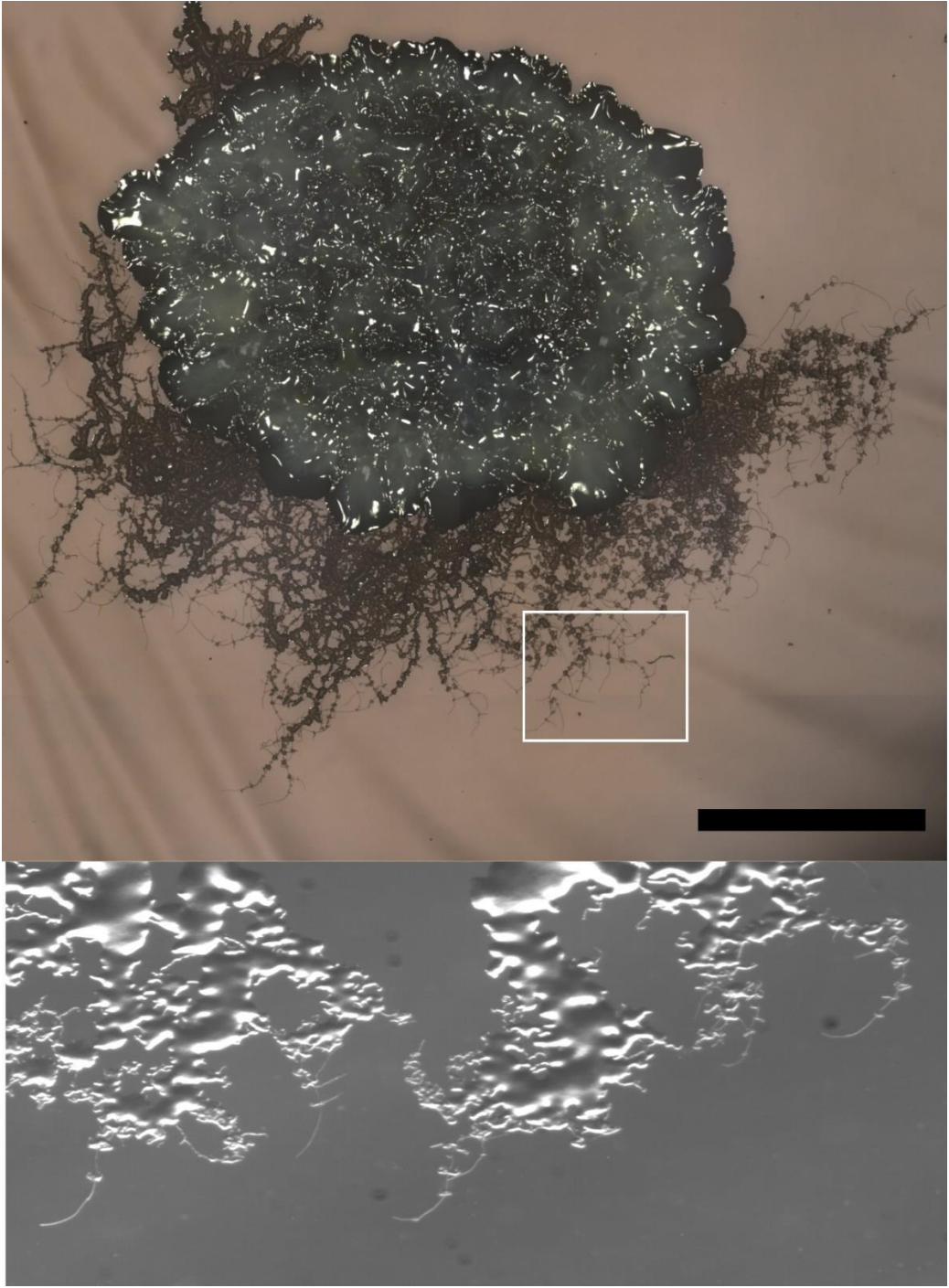

***Supplementary Fig. 14.*** Digital microscope image of environmental sample. Chiral biofilm forming bacteria escape from crowded region by forming CW rotating chiral branches at room temperature. (Bottom), magnified image of the chiral branches.